\documentstyle[12pt,epsfig,rotate]{article}

\begin{document}

\newcommand{\tb}{${\rm tan}\beta$}
\newcommand{\s}{\smallskip }

\def\thefootnote{\fnsymbol{footnote}}
\setcounter{footnote}{0}

\vspace{1cm}

\hfill LPT--ORSAY--11/59

\hfill CERN--PH--TH/2011--157

\vspace*{1.5cm}

\begin{center}

{\large\bf Clarifications on the impact of theoretical uncertainties on the}

\vspace*{1mm}

{\large\bf Tevatron Higgs exclusion limits\footnote{Extended version of talks
given at several winter conferences by the authors.}}

\vspace{1cm}

{\large Julien Baglio$^{1}$, Abdelhak Djouadi$^{1,2}$ and Rohini Godbole$^{3,4}$} 

\vspace*{5mm}

{\small $^1$ Laboratoire de Physique Th\'eorique, Universit\'e Paris-Sud XI et CNRS,
F-91405 Orsay, France.\\
$^2$ Theory Unit, CERN, 1211  Gen\`eve 23, Switzerland.\\
$^{3}$ Center for High Energy Physics, Indian Institute of Science, Bangalore 
560 012, India.\\
$^4$ Inst. for Th. Physics and Spinoza Inst., Utrecht Uni., 3508 TD Utrecht, The Netherlands.
}

\end{center}

\vspace{1.5cm}

\begin{abstract}  In this note, we respond to the comments and criticisms made
by the representatives of the CDF and D0 collaborations on our recent papers  in
which we point out that the theoretical uncertainties in the Higgs production
cross section have been largely underestimated and, if properly taken into
account, will significantly loosen the Tevatron Higgs exclusion bounds. We 
show  that our approach to the theoretical uncertainties  is reasonable and
fully justified. In  particular, we show that our procedure is not very
different from that adopted by the LHC experiments and if the latter is used in
the Tevatron case, one obtains much larger uncertainties that those assumed by
the CDF and D0 collaborations.  Furthermore, we provide additional details on
our statistical analysis of the CDF and D0 exclusion  limit and show that it is
conceptually correct. 

\end{abstract}

\def\thefootnote{\arabic{footnote}}
\setcounter{footnote}{0}

\newpage

\subsection*{1. Introduction}

In two earlier papers \cite{JHEP,PLB}, we updated the theoretical predictions
for the production cross sections of the Standard Model Higgs boson at the 
Tevatron collider, focusing on the main search channel, the gluon--gluon fusion
mechanism $gg\to H$ \cite{ggH-LO}, including the relevant higher order QCD
[4--9] 
and electroweak corrections \cite{ggH-radja,ggH-EW}.  We then estimated the
various theoretical uncertainties affecting these predictions: the scale
uncertainties which are viewed as a measure of the unknown higher order effects,
the uncertainties from the parton distribution functions (PDFs) and the related
errors on the strong coupling constant $\alpha_s$, as well as the uncertainties
due to the use of an effective field theory (EFT) approach in the determination
of the radiative corrections in the process at next-to-next-to-leading order
(NNLO). We found that contrary to the Higgs--strahlung processes \cite{HV-all},
where the rates are well under control as  the uncertainty is less than
$\approx 10\%$, the theoretical uncertainties are rather large in the case of
the gluon--gluon fusion channel, possibly shifting the central values of the
NNLO cross sections by up to $\approx 40\%$. These uncertainties are thus
significantly larger than the $\approx 10\%$--20\% error assumed by the CDF and
D0 experiments in their  analysis that has excluded the Higgs mass range
$M_{H}\!=\!158$--175 GeV at  95\% CL \cite{Tevatron0,Tevatron1,CDF}. As $gg\to
H$ is by far the dominant Higgs  production channel in this mass range, we
concluded that the above exclusion limit should be reconsidered in the light of
these large theoretical uncertainties.\s

After our papers appeared, some criticisms have been made by the members   of
the CDF and D0 collaborations and of the Tevatron New Physics and Higgs working
group (TevNPHWG) \cite{Response1,Response2} concerning the theoretical modeling
of the $gg\to H$  production cross section that we proposed. This criticism was
made more explicit  in the Tevatron Higgs talks this winter at the La Thuile
\cite{Thuile} and Moriond--QCD  \cite{Moriond} conferences  (where a long
discussion on the theoretical uncertainties has been scheduled after the talk of
one of the authors \cite{Moriond-AD}).\s

In this note, we respond to this criticism point by point and show that  that
our approach to the theoretical uncertainties  is fully justified. In
particular, we will make use of of a recent  collective effort \cite{LHCXS} made
by theorists along with  experimentalists of the ATLAS and CMS collaborations
to evaluate the Higgs  cross section at the LHC, with a special attention to the
gluon fusion  mechanism which is also the process of interest here.  Several 
issues discussed in our papers \cite{JHEP,PLB}  have been indeed addressed  in
the report of this working group. It turns out that many of the proposals that 
we put forward for the $gg\to H$ process are in fact similar to those adopted in
this comprehensive LHC study. We will thus also use the conclusions  of this
report (together with other studies that appeared  very recently) to strengthen
some of our arguments even more.\s

Another criticism made by the CDF and D0 collaborations is on the  statistical
analysis of the exclusion  limit that we performed in Ref.~\cite{PLB}, using 
the detailed information and the multivariate analysis given in a CDF paper
\cite{CDF}. Apparently, there was  a misunderstanding on what we actually did in
our ``emulation" of the CDF/D0 limit: we did not increase the theoretical
uncertainty (or add an extra uncertainty) but simply changed the normalisation
as if the cross section was evaluated using another set of PDFs (such as 
HERAPDF \cite{PDF-HERA} or ABKM \cite{PDF-ABKM} rather than the adopted MSTW
choice \cite{PDF-MSTW,PDF-MSTW1}). In this case, using the neural network output of the
CDF analysis to re-estimate the sensitivity and the exclusion limit is fully
justified and our analysis is conceptually correct. \s

Finally,  we take this opportunity to correct an error made in Ref.~\cite{PLB}
in the numerical evaluation of the $gg\to H$ cross section using the HERAPDF
\cite{PDF-HERA} set.   This error will only slightly change part of the
discussion in Ref.~\cite{PLB} and will not alter our general conclusions.

\subsection*{2. Theoretical uncertainties on the $\mathbf{gg\to H}$ cross section}

{\bf A. The scale uncertainty}\s 

It is a known and well accepted fact that the choice for the domain of scale
variation, which is supposed to account for the missing contributions at higher
orders in perturbation theory, is subjective. This is true together  with the
fact the ``scale variation only gives a lower limit on the {\it true}
uncertainty from higher orders" (we use here the same words are those given in
Ref.~\cite{LHCXS}).  However, there are at least three arguments which make us
believe that, in the particular case of the $gg \to H$ process at the Tevatron,
the situation is really exceptional and the domain of scale variation should be
extended from  the usual choice (which, we stress again,  is only a guess and by
no means a dogma) $\frac12 \mu_0 \le \mu_R, \mu_F \le 2\mu_0$ (i.e.
$\mu_0/\kappa \le \mu_R, \mu_F \le \kappa \mu_0$ with a factor $\kappa=2)$
\cite{ggH-FG,ggH-radja}, with the central scale taken to be\footnote{The choice 
$\mu_0=\frac12 M_H$ for the central scale, instead of the (for a long time
supposedly most natural) value $\mu_0=M_H$, is motivated by the fact that it
leads to a better  convergence of the perturbative series; in addition, it
implicitly takes into account  the soft--gluon re-summation contributions which
are at the level of $+15\%$ for $\mu_0=M_H$. Note that a scale variation with 
$\kappa=2$ around the value $\mu_0=\frac12 M_H$, will bring us   back only to
the original central scale value at most, which we feel is rather
optimistic.}, $\mu_0=\frac12 M_H$ \cite{ggH-radja},  to at least the range
$\frac13 \mu_0 \le \mu_R, \mu_F \le 3\mu_0$ (i.e. with $\kappa=3$). Here,  we
stick to a discussion of  the QCD  corrections up to NNLO only; additional QCD
contributions beyond this perturbative order as well as the electroweak
corrections  will be addressed later.\s

$i)$ The K factor, i.e.  the effect of the higher order QCD corrections, in the
$gg \to H$ process at the Tevatron is extraordinarily large. It increases the LO
cross section by a factor of three and it is in fact  this factor of three which
makes the CDF/D0 experiments sensitive to the Standard Model Higgs boson with
the presently collected data. There is basically no other electroweak process 
which receives such large contributions from higher orders (as long as no new
coupling or a different type of contribution not present at LO does not appear
present in these higher orders, as in fact is the case here). If we stick only  
to NNLO, then one should really be worried about the convergence of the 
perturbative series as the K factor is 2 at NLO and 3 at NNLO. A factor 
$\kappa=2$ for scale variation to estimate the higher orders would thus be 
justified for any process where the QCD corrections are moderate (like  most 
other processes that have been discussed so far) but not in this special 
case\footnote{We note that the original analyses of the  higher order
corrections to $gg\to H$  had mainly considered the case of  the LHC (the
extension for the Tevatron was made for completeness as it was not clear if the
process was viable there) and, in this case, we fully support the choice of
$\kappa=2$ as the  QCD corrections lead to only a total K factor of two with an 
acceptable convergence, as it goes from $\approx1.7$ at NLO to   $\approx2$ at
NNLO.}. We note that the  scale variation of the NLO cross section barely
reaches the central value of the cross section at NNLO for $\kappa=2$ (not to
mention  the LO cross section which needs a factor $\kappa=4$ to contain the
NNLO  central value); to have  a {\it significant} overlap of the NLO scale
uncertainty band with the NNLO central value, the choice $\kappa=3$ is more
appropriate\footnote{Note that in  Ref.~\cite{LHCXS}, the choice $\kappa=3$ has
been adopted for the Higgs--strahlung processes $pp\to HW/HZ$ despite of the
fact that the QCD corrections are much smaller than in the $gg$ fusion
channel.}. \s

$ii)$ Another argument to increase the domain of scale variation from $\kappa
=2$, which leads to an uncertainty of $+10\%, -12\%$ for say $M_H=160$ GeV, to
$\kappa=3$ which gives an uncertainty of $+15\%,-20\%$ for the same $M_H$ value,
has been  given in the addendum to Ref.~\cite{JHEP} and is reproduced below.  If
the NNLO $gg\to H$ cross section  is broken into the three pieces with 0, 1  and
2 jets, and one applies a scale variation for the individual pieces in the 
range $\frac12 \mu_0 \leq \mu_R, \mu_F \leq 2 \mu_0$, one obtains with 
selection cuts similar to those adopted by the CDF/D0 collaborations, a scale
uncertainty on the ``inclusive"  cross section that is about $+20\%,-17\%$, when
one averages over the various final states with their  corresponding weights
\cite{ADGSW}. This is very close to the result obtained in the CDF/D0 analysis 
\cite{Tevatron1} which quotes a scale uncertainty of $\approx \pm 17.5\%$ on the
total cross section, when the weighted uncertainties for the various jet cross
sections are added.  Thus, our supposedly ``conservative" choice  $\frac13 \mu_0
\leq \mu_R=\mu_F \leq 3 \mu_0$ for the scale variation of the total inclusive
cross section $\sigma^{\rm NNLO}_{\rm gg\to H}$  leads to  a scale uncertainty
that is very close to that obtained when one adds the scale uncertainties of the
various jet cross sections for a variation around the more ``consensual" range
$\frac12 \mu_0 \leq  \mu_R, \mu_F \leq 2 \mu_0$. We also note that when breaking
$\sigma^{\rm NNLO}_{\rm gg\to H}$ into jet cross sections, an additional error
due to the acceptance of jets is introduced; the CDF and D0 collaborations,
after weighting, have estimated it to be $\pm 7.5\%$. This error, combined with
the weighted uncertainty for scale variation, will certainly increase the total
scale error in the CDF/D0  analysis, possibly (and depending on how the errors
should be added) to the level where it almost reaches or even exceeds our own
supposedly ``conservative" estimate. \s

$iii)$ The above issue brings us to a last argument which appeared only very
recently. In an analysis of the Higgs+0 jet cross section (which is the 
topology to which CDF and D0 are by far most sensitive), the authors of
Ref.~\cite{Scale-H0j} show that imposing a tight jet veto to select this
topology induces large double logarithms which significantly modify the Higgs
production cross section. They calculate the Higgs+0 jet cross section from
gluon fusion at  next-to-next-to-leading-logarithmic (NNLL) order, fully
incorporating the NNLO fixed order results and their conclusion is as follows:
``At this order  (NNLL), the scale uncertainty is 15--20\%, depending on the
cut, implying that a larger scale uncertainty should be used in current Tevatron
bounds on the Higgs". This is a major issue which has to be addressed  by 
CDF/D0   (also at the LHC where indeed an uncertainty of 15--20\% is adopted
\cite{LHC-jet})  and before it is settled, it would be perhaps wise to increase
the scale uncertainty to a level that is even larger than the one we are 
advocating. However, clearly we do not so at present.\s

Let us now briefly discuss the higher order corrections beyond NNLO to the
$gg\to H$ cross section, and justify why we do not take them into account  (as
is also done by the CDF/D0 and the LHC \cite{LHCXS} collaborations in their
analyses): 

-- As mentioned above, in the actual  CDF/D0 analysis \cite{Tevatron1,CDF}, the
$gg\to H$ cross section has been broken into the three pieces pieces
corresponding to the  production of a Higgs with 0, 1  and 2 jets and only the
NNLO QCD corrections to these jet cross sections have been taken into  account.
One should therefore stick, for consistency,  to an inclusive cross section
(which is the sum of these jet cross sections) at the same order, i.e. NNLO. 

-- For the   consistency of the calculation, the corrections beyond NNLO
(including soft--gluon resumation), need to be folded with PDFs that are at the
same perturbative order. No PDF beyond NNLO is available at the moment and,
thus, one should stick to NNLO. 

-- The soft--gluon resumation and the higher order effects beyond NNLO address
especially the logarithmic corrections (as also do the scale uncertainties).
However, it is well known that in the $gg\to H$ process, very large
contributions are coming from  constant, in particular $\pi^2$, terms. These
terms are not dealt with by the calculations beyond NNLO and their inclusion
might in turn  induce large corrections that  are not accounted for by scale
variation. 

-- An attempt  to only partly re-sum these large $\pi^2$ terms has been made in
Ref.~\cite{ggH-Ahrens}.  This is the paper to which Refs.~\cite{Moriond,Thuile}
refer to as the study in which  smaller uncertainties than those that are
advocated by all analyses are claimed. For a discussion of this analysis,  we
simply refer to the report of the experts that contributed to the the Higgs
cross section working group at the LHC (section 2.5 of Ref.~\cite{LHCXS}) in
which  arguments have been  presented to show why the uncertainties advocated in
this paper have been underestimated and reasons not to adopt them in the report
\cite{LHCXS} have been given. \bigskip

{\bf B. Uncertainties from the use of an Effective Field Theory (EFT) approach}
\s

The uncertainties from the use of the EFT approach originate  from three
different sources and we summarize their impact below.\s  

$i)$  There are first the uncertainties from the exact NLO electroweak
corrections and whether these corrections  should be included in the complete or
in the partial factorization approaches. This has been discussed in  detail in
Ref.~\cite{ggH-EW} (where the full NLO electroweak correction itself  has been
derived) and it was advocated that an uncertainty which is the difference
between the results  obtained in the two  approaches should be included. In our
paper, we have simply followed this recommendation which, for a Higgs mass of
160 GeV, leads to an  uncertainty of about  3\%. For the three--loop mixed
QCD--electroweak corrections (that, incidentally, are also included in our
calculation) there is an additional  problem: they have been calculated in the
EFT approach in which $M_H \ll M_W$ which is obviously not a valid limit.  These
corrections should therefore be taken with care\footnote{Indeed, if the NLO
electroweak corrections have been calculated in the same limit, as was done in 
Ref.~\cite{Paolo},  one would obtain a result which is completely different: 
for instance a correction of less than 0.2\% is obtained for the leading $m_t^2$
correction to the $gg \to H$ cross section \cite{Paolo} in the EFT approach $M_H
\ll M_t$, compared to several percent for the electroweak correction  in the 
exact case (another example is the two loop electroweak correction to the $\rho$
parameter where the leading $m_t^4$ correction in the EFT approach 
\cite{Hoogeven} is completely different from the result obtained with a more
refined calculation as, for instance, in Ref.~\cite{Awramik}).}. Nevertheless,
first, we included this correction as stated before and second, we did not
assign an uncertainty to this correction and sticked to the one advocated in
Ref.~\cite{ggH-EW}, i.e. included only the difference between the complete and
partial factorization approaches. \s

$ii)$ The renormalisation scheme dependence for the $b$--quark mass in the
$b$--quark loop contribution to the $ggH$ amplitude. Here, we indeed find a
1--2\%  uncertainty which is more or less equivalent to what has been  discussed
in Ref.~\cite{ggH-radja}, but which {\it was not included} in the final numbers
for the total uncertainty on the $gg\to H$ cross section, as is discussed below.
\s

$ii)$ The EFT approach for the NNLO QCD contribution:  we fully agreed with the
experts that the infinite top mass limit is very good for the  Higgs masses that
are relevant at the Tevatron (but not at the LHC for $M_H \geq 2m_t$), so there
is no problem here.  The problem is the $b$--quark loop contribution for which the
EFT approach is certainly not valid and, for instance, the omission of this
contribution at LO leads to a 10 \% difference compared to the exact case.
Furthermore, the NLO K factor for the $b$--quark loop, which is about
$K=1.2$--1.4, is much smaller than that for the top--quark loop. In fact, the
problem arises mainly because of the significant negative interference  between
the top and bottom loop contributions (the bottom quark contribution itself is
rather small) up to NLO and the fact that the bottom contribution has been {\it
factorized out} in the LO cross section. Thus, including the top quark
contribution only using the EFT approach might overestimate the NNLO correction
as this interference component is missing (and, since it is resulting simply
from an approximation in the calculation it is not, in principle, accounted for
by  e.g. the scale variation that is performed at NNLO). We have estimated this
uncertainty on the base of what is known at NLO (where both the exact and EFT
results are available \cite{SDGZ}) and taking into account the relative K
factors for the $t$ and $b$ loops at this order and we obtain an uncertainty
of a few percent. This uncertainty has not been discussed  elsewhere and we
believe that it should be definitely included as it is  not accounted for by
anything else. \s 

Contrary to what is stated in Ref.~\cite{Thuile}, {\it none} of these
uncertainties have been included  in the analysis of Ref.~\cite{ggH-radja} (nor
in those of Ref.~\cite{ggH-FG}) on which the CDF/D0 limits are based (some of
these uncertainties  have been indeed discussed in Ref.~\cite{ggH-radja} but not
included in the final numbers that have been given).  In fact, all these
uncertainties have been addressed in the LHC Higgs cross section report
\cite{LHCXS} (in which all the authors of Refs.~\cite{ggH-radja,ggH-FG} have
contributed) and we summarize their effect (see section 2.3 of \cite{LHCXS}): 
1\% uncertainty for  the electroweak contribution, 1 to 2\% from the $b$--quark
mass and  1\% uncertainty  from the EFT approach to the top quark contribution
(the top--bottom quark loop interference has not been taken into account). This
makes a total of up to 4\% uncertainty which is not far from our 5\% estimate
(note that, unfortunately, these uncertainties have also not been included in
the final numbers given in Ref.~\cite{LHCXS}, but  we have been told that it
will be the case in an update of the analysis that will appear soon). We stress
that these are  only estimates and some arbitrariness will remain until further
higher order or exact contributions are calculated.  \bigskip


{\bf C. The PDF uncertainties}\s

Concerning the PDF uncertainties, there are two issues on which there is 
criticism: first, the way we estimate  the uncertainties within the MSTW set of
PDFs that is usually adopted  and second, our later use of the  ABKM09 and 
HERAPDF parameterizations to illustrate the impact of the   PDF uncertainties on
the $gg\to H$ cross section at the Tevatron.  We will present a detailed 
discussion of the last issue in the next section of this note. \s

Let us begin by stressing the fact that our first and main approach was to use
the (commonly adopted) MSTW2008 set of PDFs \cite{PDF-MSTW,PDF-MSTW1} and their
associated  uncertainties.  But first, we used the 90\% CL combined
PDF+$\Delta^{\rm exp}\alpha_s$ uncertainty  and, second, we added  to that  (in
quadrature) an estimate of the theoretical uncertainty in $\alpha_s$ 
($\Delta^{\rm th} \alpha_s=0.002$ at NNLO, given by the MSTW collaboration
itself but not used in the final analysis) and obtained, for $M_H=160$ GeV,  a
total of $\pm 15\%$  PDF+$\alpha_s$ uncertainty as stated in the paper. This is
exactly what one obtains if the PDF4LHC recommendation  \cite{PDF4LHC} (which,
incidentally, appeared after our first analysis was published) is used. Indeed,
the PDF4LHC recommendation to estimate the uncertainties is that one takes the
envelope of the PDF+$\Delta^{\rm exp}\alpha_s$ uncertainties  obtained using the
three sets given by the MSTW, CTEQ \cite{PDF-CTEQ} and NNPDF \cite{PDF-NN}
collaborations. For the  $gg\to H$ cross section at NNLO,  to a very good
approximation, this reduces to taking  the PDF+$\Delta^{\rm exp}\alpha_s$ MSTW
uncertainty at the 68\% CL  and multiply it by a factor of two. One  then
obtains  15\% PDF+$\alpha_s$ uncertainty for $M_H= 160$ GeV, i.e. {\it almost
exactly} the value that  we assume in the paper. So we are  arriving, using two
different ways, to the same result. \s

We nevertheless believe that the PDF4LHC recommendation, when taken literally
(i.e. when using the prescription above), is not sufficient to account for all
possible sources of uncertainties. First, the CTEQ and NNPDF groups have only
parameterizations at NLO, while we are evaluating our process at NNLO and thus,
only one PDF set can be used consistently in this particular case.  Second, it
would have been rather unfair to ignore in the final recipe to calculate the
uncertainty the other PDF sets  which, instead,  are at NNLO. And indeed, 
contrary to statements that we heard, the PDF4LHC does not recommend to ignore
totally the other PDF sets.   The complete and correct statement of the  PDF4LHC
group is (see end section 2.2 of Ref.~\cite{PDF4LHC}): ``Since there are NNLO
PDFs obtained from fits by the ABM, GJR   and HERAPDF groups, these should
ideally be compared with the above procedure". This is exactly what we did in
our papers. \s

Thus, we end this discussion of PDF related uncertainties by  noting that these 
uncertainties seem to been underestimated in the CDF/D0 analysis  (and, in fact,
also  in our original paper Ref.~\cite{JHEP}). The second issue of  consistency
of the predictions of the ABKM and HERAPDF  sets  with the Tevatron data  and
that of the large difference in the gluon densities between the various PDF sets
on the other hand is deferred to section 3.\s

{\bf D. The combined scale+PDF uncertainty}\s

Finally, we address  the issue of combining the uncertainties which, as we
stated in our answer to Ref.~\cite{Response2} in the addendum to
Ref.~\cite{JHEP}, is the only relevant issue in this context and it explains the
major part of the difference  between the total uncertainties that we assume in
our paper compared to that adopted by the CDF collaboration.\s

Let us start by an important comment: we do not add scale+EFT and the PDF
uncertainties linearly as often stated. Our procedure for the 
combination has been presented in Ref.~\cite{JHEP} and is the following:  we
calculate the maximal and minimal cross sections with respect to the scale
variation and apply on these cross sections the PDF+$\Delta^{\rm exp} 
\alpha_{s}$+$\Delta^{\rm th} \alpha_{s}$ uncertainty.  This procedure has been
in fact already proposed, together with other possibilities which give similar
results, in Ref.~\cite{PDF-ttpaper} where  top quark pair production at hadron
colliders was discussed.  To this combined scale+PDF uncertainty, one can
linearly add the small scheme/EFT uncertainty  to obtain the overall theoretical
error. Nevertheless, it turns out that our procedure for adding the scale and
PDF uncertainties leads to a total uncertainty that is close but a little bit 
smaller than what we would get had one added them linearly.  \s

Note that our  procedure above would eventually take into account possible 
correlations between the scale and the PDFs (which are evaluated at a given
factorization scale).  Indeed, an argument for our approach  put forward in 
Ref.~\cite{PDF-ttpaper}  is that unknown high order effects also enter in the
PDF determination and one needs to use the full information  on the PDF
uncertainties in the determination of the scale dependence. Nevertheless,  we
admit that the correlations between the PDF and the scales might be indeed small
as stated in Ref.~\cite{Response2}. We believe, however,  that the effects of
this correlation  cannot be reliably estimated at present;  see also the
discussion at the end of section 12.5.2 in Ref.~\cite{LHCXS}.\s

Another important aspect is that the uncertainties in the PDFs discussed up to
now are only those due to the errors in the data included in the fits and are
thus of experimental  origin (and indeed have a probabilistic meaning). However,
there are many other possible sources of uncertainties which are, this time, of
theoretical and not  experimental nature: scale variation and higher order
effects in the observables  used in the fits, ambiguities in heavy quark flavor
scheme definition, difference between the $\alpha_s$ values that one obtains
from DIS and LEP data, etc...  These theoretical PDF uncertainties cannot be
estimated within a given  parametrisation, say the MSTW parametrisation. The
only reasonable way to estimate them is to compare the predictions for the
central values of the cross section given by different parameterizations (which
mostly use the same data). The PDF theoretical uncertainties would be then
equivalent to the spread that one observes when comparing different PDF sets. In
this case, the PDF uncertainties should be considered as having no statistical
ground and would for instance,  simply change the normalisation of the cross
section.\s  

In our approach when dealing with the MSTW parametrisation only, since  we
cannot determine the pure theoretical uncertainties, we have simply interpreted 
the Hessian errors as a measure of the  uncertainties due to the theoretical
assumptions in the parametrization of the quark  and gluon densities (and, as
stated before,  equivalent to the spread that one observes when comparing
different parameterizations). In this case,  the PDF uncertainties  cannot be
combined in quadrature with the two other theoretical errors, namely the scale
and the scheme/EFT uncertainties: all should be considered as pure theoretical
uncertainties with a flat prior and added linearly.\s 

In fact, even if the PDF+$\alpha_s$ uncertainties are taken from the Hessian
method and considered as experimental with  a Gaussian behavior, it is still not
obvious  that they should be added quadratically to the scale uncertainties
which have a flat distribution being purely theoretical. The (not obvious)
procedure  of combining a Gaussian and a flat distribution has been addressed 
by the LHC Higgs cross section working group (see section 12 of
Ref.~\cite{LHCXS}) and the conclusion was: ``As a general rule that is
sufficiently conservative only the linear combination of these  (scale and PDF)
errors can be recommended" (next to last paragraph of the concluding section
12.5.2). Thus, contrary to the statements of the CDF/D0 collaborations  there is
no consensus on how to add a flat and a Gaussian distribution.\s

In conclusion, we believe that our approach to combine the theoretical
uncertainties is fully justified and, in fact, not the most conservative one
(see the next point for a more radical but still justified possibility). If  we
follow our approach, the combined scale, EFT and PDF+$\alpha_s$ uncertainties
will lead to an $\approx +41\%,-37\%$ total uncertainty on the $gg\to H$ cross
section at the Tevatron in the mass range $M_H \! \approx \! 160$ GeV with
almost the best sensitivity. This is to be compared to the $\approx\! 10\%$ 
and  $\approx\! 20\%$ uncertainties assumed, respectively, by the D0 and CDF
collaborations. For comparison, the result obtained when one adds linearly,
i.e.  as recommended by the LHC Higgs cross section working group,  the 
uncertainties from scale ($\!+\!20\%,\!-\!17\%$ on the sum of the jet cross
sections) and PDFs ($+16\%,-15\%$ when the MSTW 68\%CL PDF+$\Delta^{\rm exp}
\alpha_s$  error is multiplied by a factor of two following the PDF4LHC
recommendation) assumed by CDF, leading to a total of $\approx \!+\!36\%,
\!-\!32\%$ for $M_H\! \approx\! 160$ GeV.\s

 Thus, the uncertainty that we assume is comparable to the one obtained when the
CDF scale and PDF uncertainties  are combined using the LHC procedure
\cite{LHCXS}, the difference being simply due to the  additional ${\cal O}(5\%)$
uncertainty from the use of the EFT approach that we also include. See also
Ref.~\cite{Sally} for an independent discussion.\s

To highlight this very important point (which wed hope will close this
discussion), we have modified the original  Fig.~2 of  our recent most paper
\cite{PLB} to include the uncertainty as calculated using the LHC procedure; we
append this figure below.

\begin{figure}[!h]
\begin{center}
\epsfig{file=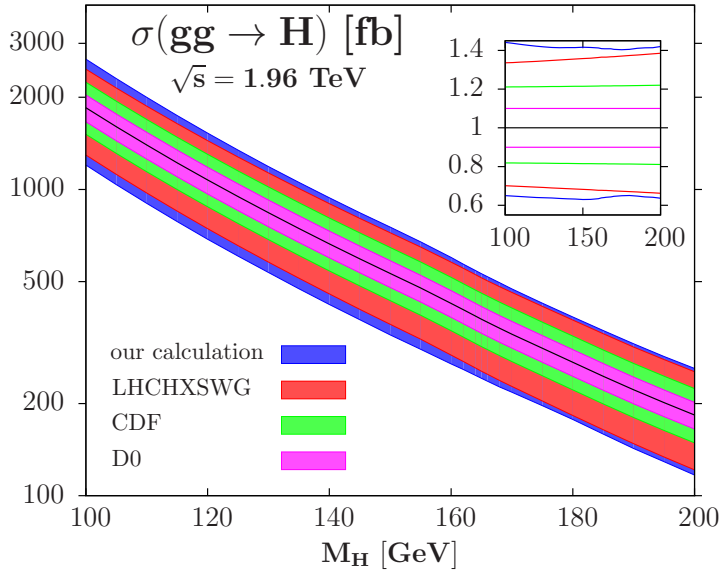,scale=0.89} 
\end{center}
\vspace*{-5mm}
\caption[]{The production cross section $\sigma^{\rm NNLO}_{gg\to H}$  at the 
Tevatron using the MSTW PDFs, with the uncertainty band when all theoretical
uncertainties are added as in Ref.~\cite{JHEP}. It is compared 
the uncertainties quoted by the CDF and D0 experiments  as well
as the uncertainty when the LHC procedure  \cite{LHCXS} is adopted.  In the
insert, the relative size of the   uncertainties compared to the central value
are shown.}
\vspace*{-3mm}
\label{Total}
\end{figure}

\subsection*{3. The HERAPDF and ABKM PDF parameterizations}

An important criticism made by the CDF/D0 collaborations concerns our choice of
considering also the impact of other PDF sets.  Let us say here at the outset
that we are only ``users" of the PDFs provided by different PDF groups. We do
not wish to advocate any one  determination to be better or worse. We simply
wished to answer a simple question as to what would be the effect on the Higgs
exclusion limit presented by the CDF/D0 collaborations, if one uses other PDF
sets than MSTW.  Please note that these  issues and differences will be
reasonably irrelevant in case of discovery, as there will be many other
observations which one can then use to clarify the situation. The whole issue of
assessing the theoretical uncertainty in the rates takes a special meaning when
we talk of possible exclusion and associate a certain statistical significance
with this exclusion. Given the fact that the experiments have not been able to
give a limit on the Higgs production cross section multiplied by the branching
ratio at a  given level of statistical significance, the translation of the
reduction in the normalisation on possible exclusion becomes a  nontrivial task.
We consider that our publication gives at least an indication of the answer,
which many a theorists would like to have, independently of the recommendation
by  the PDF4LHC group. \s

With respect to the comments in Refs.~\cite{Thuile,Moriond}  regarding the use
of HERAPDF and ABKM sets, we summarize different relevant facts which
address those comments below.\s 

$i)$ An understanding of the difference in the gluon densities at NNLO obtained
by  MSTW on the one hand and ABKM/HERAPDF on the other, is still a matter under
investigation and being studied by experts. For instance, the last word is 
still to be said about the recent ABM analysis of the possible large  effect of
the NMC data on the NNLO parameterizations \cite{NMC}.  This analysis shows that
the constraints from fixed--target DIS data, in particular data from the NMC
experiment, in the  PDF fits at NNLO (the impact at NLO is expected to be small
so that CTEQ and NNPDF might not be affected) can play a very important role in
explaining the large differences between MSTW on the one hand and ABKM/HERAPDF
on the other hand, both  in the extracted values of the $\alpha_s$ coupling and
of the gluon densities at high Bjorken--$x$.   Both these  effects  could impact
the  Tevatron Higgs exclusion limits and, for instance,  the conclusion of the
ABM paper states: ``[...] the current range of excluded Higgs masses at the
Tevatron appears to be much too large". A final word is still to be said and
perhaps it is too early to draw definite  conclusions from this new analysis
(which the community is looking at very seriously  as the consequences for
Tevatron and LHC can be  indeed far reaching), but this opens up the possibility
that some effects   which have now become important due to the increased
precision of the theoretical calculations might have been overlooked in some of
the analyses of the PDF determinations and hence the uncertainties might
have  been underestimated.  Since we are not experts in the subject, we only
take the observations of Ref.~\cite{NMC} to mean that the situation is not clear
in this respect.\s

$ii$) We also have to realise that now with increasing accuracy of the
theoretical calculations and data, the way certain corrections, such as  target
mass  corrections to the Deuterium data, possible nuclear effects, were
incorporated  in the global fits have had to be revisited by the groups which
make global  fits and they have also observed certain tensions. \s

$iii)$ It was mentioned in Refs.~\cite{Thuile,Moriond}  that the ABKM09
parametrization \cite{PDF-ABKM} should not be adopted  as it uses only the DIS
data in the determination of the gluon densities and that it does not give an
acceptable fit of the Tevatron jet data. This is not entirely correct.  Indeed,
Alekhin, Blumlein and Moch provided very recently a new PDF set \cite{ABM10}, in
which the D0 Run II dijet data are included. It turns out that the impact of
these data on the ABKM fits is only marginal and  that ABKM09 provides indeed a
very good description of the Run II jet data. In  fact this parametrization
seems to perform satisfactorily~\cite{PDF-all} and for a detailed discussion, 
we refer to the two talks quoted in Ref.~\cite{ABM10}.\s

$iv)$  It is true that HERAPDF does not use any jet (either from the Tevatron or
from their own DIS experiments) data so far in their determination of parton
densities. The ZEUS experiment has demonstrated that the use of jets produced in
the DIS process will in fact reduce the uncertainty in the knowledge of gluon
density, but as things stand this has not been included so far. The interesting
and special feature of the HERAPDF group is that they determine the flavour 
decomposition using {\it only} HERA data, without using the $\nu$-DIS data or
the lower energy charged lepton DIS data, most of which is  with nuclear
targets. This makes them qualitatively different from the other data which use
global fits. Hence, the predictions with HERAPDF for processes at the Tevatron 
or the LHC should be taken into account while studying the spread of the 
predictions due to the PDF uncertainties as this gives a better  estimate of the
uncertainties or our ignorance. In Moriond--QCD, a talk given by Voica Radescu
\cite{Voica} shows that HERAPDF performs very well in describing the Tevatron
data. This is particularly true for the $W/Z$ data which are described at NLO
and, thus, have the gluon splitting there and give an indirect test of the gluon
density.\s

The above  points are made only to say that there are issues about the PDF  fits
that need more investigation and till the dust has settled on this, one should 
use ABKM and HERAPDF predictions as a reflection of the  theoretical 
uncertainty in the game.  As mentioned already, since apart from MSTW, ABKM and
HERAPDF are the only parameterizations (together with  GJR) which are
available\footnote{The NNPDF collaboration is also releasing a PDF set a NNLO.}
at NNLO, it becomes imperative  that one assesses the impact of using these for
the Higgs production cross section. Incidentally and as mentioned earlier, even
the PDF4LHC recommendation  asks  that one should  study the effect of using
other parton densities. This is particularly important, for a  crucial issue
such as the exclusion of the Higgs boson in a certain mass range. In fact, by
showing that the use of other PDF sets can have such a dramatic consequence for
the (non)exclusion of the Higgs boson at the Tevatron, we had hoped that one
could urge the community to understand the difference between the global  fits
and only the DIS fits, urgently.\s

\begin{figure}[!h]
\begin{center}
\vspace*{1mm}
\epsfig{file=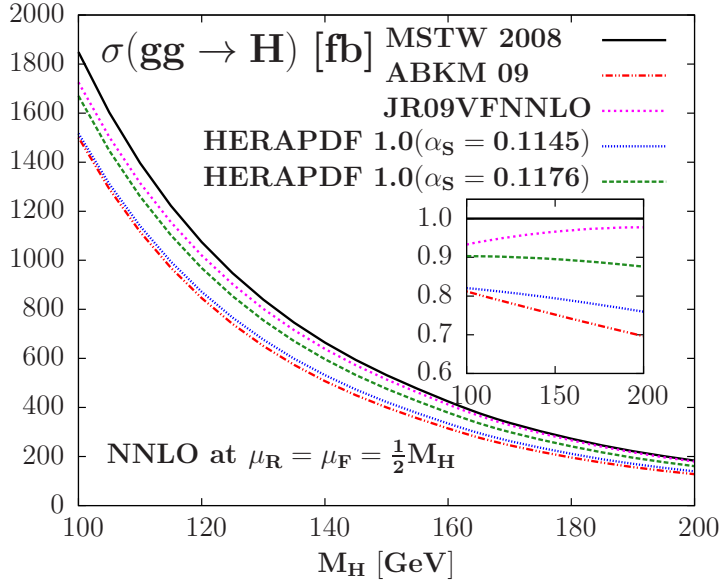,scale=0.89} 
\end{center}
\vspace*{-5mm}
\caption[]{The $gg\to H$ cross section as a function of $M_H$ when the four 
NNLO PDF sets, MSTW, ABKM, JR and HERAPDF, are used. In the inserts, shown
are the deviations with respect to the central MSTW value.}
\vspace*{-2mm}
\label{PDFs-correct}
\end{figure}

Finally, let us take this opportunity to make a {\it mea culpa}.  After our
paper \cite{PLB} had appeared, we realized that an error occurred in the
numerical analysis which had led to Fig.~1 of Ref.~\cite{PLB} for the $gg\to H$
production cross section when the four NNLO PDF sets are adopted. In the plot 
with  the  two HERAPDF sets, the central scales at which $\sigma^{\rm
NNLO}_{gg\to H}$ has been evaluated were not set to $\mu_R=\mu_F=\frac12 M_H$ as
it should  have been, but at $\mu_R=\mu_F=\frac32 M_H$ which gives the minimal
cross section once the scale uncertainty is included. This explains the large
difference in the cross  section\footnote{We thank Graham Watt for pointing out
to us that his calculation of the $gg\to H$  cross section with HERAPDF  does
not lead to such a large difference.}, up to 40\%, between the MSTW and HERAPDF
predictions\footnote{ Note that the same analysis presented for the LHC in 
Ref.~\cite{JHEP-LHC} is not affected by this problem.} .  We thus present our 
apologies and produce in  Fig.~\ref{PDFs-correct} the correct  figure where all
scales are  consistently set to $\mu_R=\mu_F=\frac12 M_H$. The difference
between the MSTW and HERAPDF  predictions reduces now to $\approx 20\%$ at most,
which is indeed much more reasonable. In this case, the smallest value of the
cross section is given  when using the ABKM set and amounts to $\approx
20\%$--30\% in the considered Higgs  mass range as noticed in Ref.~\cite{JHEP}
(this difference is slightly larger if the new ABM10 PDF set  is used and for
$M_H\approx 160$ GeV, one has a $\approx 30\%$ difference \cite{PDF-all}).\s

This error does not affect the subsequent discussion and almost does not change
our conclusions. Indeed, the main analysis which led to Fig.~1 (which, we
believe,  is the most important result of our papers) is still valid as we
estimate the PDF uncertainties within the MSTW set and our conclusion, that  the
theoretical uncertainty on the $gg\to H$  cross section at the Tevatron is
$\approx 40\%$, still holds true.

\subsection*{4. Emulation of the Tevatron limit calculation}

We come now to the specific issue of our emulation of the CDF sensitivity on the
Higgs boson that we have performed in Ref.~\cite{PLB}. Let us first summarize
our main result: by simply adopting a different choice of the PDF set for the
signal and main background cross sections, we arrived at a needed  luminosity 
which should be approximately  a  factor of two larger than the one assumed in 
the CDF paper (i.e. 5.9 fb$^{-1}$) in order to recover the present sensitivity.
A few important remarks can be made on our statistical treatment.\s 

-- First of all, one possible way to estimate the uncertainties related to the
PDFs is to take the spread in the different predictions of the cross sections
when evaluated with  the various sets of NNLO PDFs  as a measure of this
uncertainty (this was how PDF uncertainties were estimated before the advent of
the Hessian method).  Since the maximal cross section for $gg\to H$ is obtained
with the MSTW set and the minimal one was thought to be with  the HERAPDF set
which gave in our (incorrect) calculation a 40\% lower rate,  the PDF
uncertainty when one takes the central value from MSTW will be $+0\%, -40\%$. 
If we now chose the parametrisation which gives the smallest $gg\to H$ cross
section as a reference set for the normalisation\footnote{ It was  apparently
understood by the CDF/D0 collaborations that  we increased our PDF uncertainty
which resulted in a larger theoretical uncertainty band and we considered the
lower limit of this band to (re)estimate the sensitivity. This is not what we
did in our paper: we are definitely not adding an extra systematic uncertainty
but simply changing the normalisation of the cross section.}, one can ignore the
PDF uncertainty  when setting the Higgs exclusion limit and the only
uncertainties which remain are the scale+EFT uncertainties, which are at the
minimum level of 20\%. One can then simply use the CDF analysis to derive the
Higgs exclusion limit but assuming a normalisation for $\sigma^{\rm NNLO}_{gg\to
H}$ that  is lower  compared to that used in the original study but with the
same 20\% uncertainty. Here, we assume indeed that only the normalisation of the
rate  will change and that  the PDF effects on kinematical distributions will
remain the same or small (this is, to our view, not a very bad approximation and
we cannot, of course, estimate ourselves the impact on the experimental
analysis). \s

-- Second, we do not pretend to have the cleanest and most complete treatment of
systematic uncertainties\footnote{These uncertainties are of course included in
our analysis since we have used the multivariate output plots of the paper that
have been obtained incorporating the full set of uncertainties.} and their
correlations, which is simply beyond the  scope  of our paper and is definitely
the job of the experimental  collaborations.  Our goal in Ref.~\cite{PLB}  is
different and more modest: it is to estimate the {\it  relative} impact on
sensitivity with the variation of  the Higgs cross section  that is  obtained
when a different choice of PDFs is made.   The estimated value  of the
sensitivity itself is not very important; what is most  important  is its
relative variation due to  the different choice of PDFs.  Thus, our estimate of
the sensitivity is simply a starting point from which the relative variation has
been evaluated. We made the effort of having this starting point as close as
possible to the CDF sensitivity. The  variation of the sensitivity is then
evaluated relatively {\it to our own estimate} of the sensitivity  and not
relatively to the one of CDF.  Therefore, the up to 30\% difference in
sensitivity between our estimate and the CDF one (in passing,  it is less than
10\% for the observed and less than 30\% for expected) should have no impact on 
the relative variation of the sensitivity and the obtained  necessary luminosity
to recover the initial sensitivity.  Note that CDF has used in the sensitivity
estimate a Bayesian method of ratio of profile likelihood, while we use instead
a simple frequentist ratio of log likelihood ``\`a la LEP"; this is a possible
source which could explain part of the difference in our results. \s

-- A final remark concerns the impact of systematic uncertainties and their
correlations in estimating the relative variation of the sensitivity.   A clean
treatment of systematic uncertainties with correlations is indeed very important
for a realistic estimate of the sensitivity and we have no doubt that this
statistical treatment has been made in  a complete way in the  CDF analysis.
However, since we are using the multivariate output plots of the CDF analysis to
estimate the sensitivities and these output plots are supposed to include the
entire information on the uncertainties and their correlations, we believe  that
our treatment is fully satisfactory contrary to the (surprising) statement of
Refs.~\cite{Thuile,Moriond}. In addition, as stated above, the (presently not
large) systematic uncertainties will have no impact in evaluating the {\it
relative variation} of the sensitivity when a different choice of PDF is made. 
Therefore, the entire discussion  on including the correlations between the
uncertainties in signal and background is not relevant in this context. \s

In fact, our main result of the needed luminosity to recover the present
sensitivity can be checked with a back-of-the-envelope calculation by
considering the counting sensitivity formula $S/\sqrt{S+B}$ and by equating $S$
and $B$. One then  arrives at the very simple formula for the required
luminosity ${\cal L}'$  as a function of the initial luminosity ${\cal L}$ and
the signal cross section variation factor $f$,  ${\cal L}' = {\cal L} \ 
(1+f)/(2 f^2)$. Setting the present luminosity to ${\cal L=}5.9$ fb$^{-1}$ as 
in the CDF analysis, one obtains for a 20\% and 40\% reduction of the $gg\to H$
cross section, needed luminosities of ${\cal  L}' = 8.3~{\rm fb}^{-1}$ for
$f=0.8$ and ${\cal  L}' = 13.1~{\rm fb}^{-1}$ for $f=0.6$. These are almost
exactly the values that we obtain in our paper, respectively, 8 and 13
fb$^{-1}$. This gives us further confidence that our analysis and our  main
result are correct.\s

\begin{figure}[!h]
\vspace*{-.6cm}
\centerline{\epsfig{file=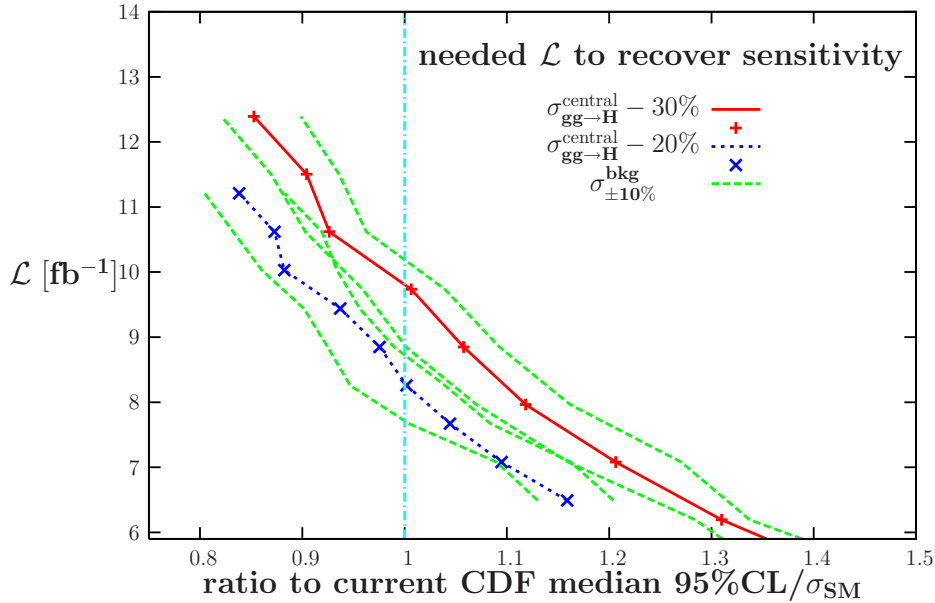,scale=0.7} }
\vspace*{-12.1cm}
\caption[]{The luminosity needed by the CDF experiment to recover the current 
sensitivity (with 5.9 fb$^{-1}$ data) when the $gg\!\to\! H\! \to \!\ell \ell 
\nu \nu$ signal rate is lowered by 20 and 30\% and with a $\pm 10\%$ change in 
the $p\bar p\! \to\! WW$  dominant background.}
\vspace*{-1mm}
\label{Lumi-new}
\end{figure}

Nevertheless, due to the error in the determination of the $gg\to H$ cross
section using the HERAPDF set,  the interpretation of  the CDF/D0 limit when
lowering the normalisation of the cross section, has to be modified. Instead of
lowering the normalisation  by 40\%, one has to lower it by 30\% which is the
difference between the  MSTW and ABKM  predictions. The luminosity needed by
the CDF experiment to recover the present sensitivity  is shown in
Fig.~\ref{Lumi-new} in this case. With this normalisation and  including the
10\% uncertainty on the background rate, the needed luminosity to recover the
present sensitivity will be slightly less than a factor of two\footnote{
Note, however, that the updated results given by the CDF/D0 experiments for the
winter 2011 conferences with a luminosity of 7.1 fb$^{-1}$ for CDF, lead to an
exclusion limit that is slightly worse than the one quoted here and only the
range $M_H=158$--173 GeV is excluded. Thus, even for a 30\% reduction of the
production cross section only instead of the 40\% used earlier, one still needs
$\approx 13$ fb$^{-1}$ data to recover the sensitivity obtained with 7.1
fb$^{-1}$.} . \s

\subsection*{4. Conclusion}

To conclude this note, let us summarize  the main points that we put forward in
our analysis of the $gg\to H$ cross section at the Tevatron.\s 

Concerning our discussion of the theoretical uncertainties in the NNLO 
production rate:\s

\begin{itemize} 
\vspace*{-2mm}  

\item[$i)$] The scale uncertainty has not been overestimated in our analysis. We gave
several arguments in favor of an extended domain for scale variation and in
fact, it turns out that our uncertainty is comparable to that assumed by the
CDF/DO collaborations when the $gg\to H$ cross section is broken into jet cross
sections and to the uncertainty advocated in Ref.~\cite{Scale-H0j} when the 
impact of the jet veto is included in the Higgs+0 jet cross section alone.
\vspace*{-2mm}  

\item[$ii)$] For the uncertainty from  the EFT approach, many of its components
have been discussed in other papers and we simply made the effort to estimate
the overall impact.\vspace*{-2mm}  

\item[$iii)$] We do not believe that we are overestimating the PDF uncertainties. In fact
the result that we quote within the MSTW set is exactly the one that is 
obtained used the PDF4LHC recommendation. In fact, we even believe that we are
underestimating these PDF uncertainties, in particular if the analysis of
Ref.~\cite{NMC} turns out to be correct.\vspace*{-2mm}  

\item[$iv)$] We do not add linearly the PDF and scale+EFT uncertainties. Our procedure,
which has been also advocated in other analyses like Ref.~\cite{PDF-ttpaper},
addressed also the theoretical part of the uncertainties. The result that we
assume is indeed close to a linear sum (in fact slightly smaller), but a linear 
combination of scale+PDF uncertainties is exactly the one recommended in the LHC
Higgs cross section  working group report \cite{LHCXS}.\vspace*{-2mm}    

\item[$v)$] If the recommendations of the LHC Higgs cross section  working group report
\cite{LHCXS} are adopted for the CDF uncertainties, one would obtain the same
uncertainties as the ones that are advocating in the paper (modulo the  small
EFT uncertainties); see again Fig.~1 that was included here.\vspace*{-2mm}  

\item[$vi)$] The various issues discussed here appear also in the case of Higgs
production in supersymmetric extensions of the Standard Model. The theoretical 
uncertainties turn out to be also rather large in the main production channels
\cite{MSSM}. 

\end{itemize}

Concerning our emulation of the CDF limit calculation:  
\begin{itemize} 
\vspace*{-2mm}  

\item[$i)$] The PDF effect is not included as being a new source of systematic
uncertainty but, rather, included as being a different choice for the PDF set
from the one adopted in the CDF analysis, and which affects only the cross
section normalisation.\vspace*{-2mm}   

\item[$ii)$] Our goal was not to re-estimate  the CDF sensitivity but the relative
variation of the sensitivity when the Higgs cross section is changed by   a
different PDF choice.\vspace*{-2mm}  

\item[$iii)$] Our results are robust regarding the systematic uncertainties and their
correlations, since we are using the multivariate outputs  of the CDF
analysis that include them.\vspace*{-2mm}    

\item[$iv)$] Our main results for the needed luminosity to recover the present 
sensitivity agree with estimates obtained in a simple and heuristic way.
We believe this agreement provides a nice check of our analysis.\vspace*{-2mm}   

\item[$v)$] It is highly desirable that the CDF and D0 collaborations provide us
with a fully cut-based analysis which will be easier to follow and reinterpret; 
we will  be more than happy if they could simply redo our analysis in
Ref.~\cite{PLB}, assume a different  normalisation of the production cross
section and reinterpret the Higgs mass limit.\vspace*{-2mm}  

\end{itemize} 

Finally, concerning the discussions on the HERAPDF and ABKM parameterizations, 
let us stress again that they provide reasonable fits to the Tevatron jet data,
contrary to an apparently common belief. There are issues about the PDF  fits
that need more investigation  (in particular the point raised recently on the
treatment of the NMC data which might lead to a significant impact) and until a
better understanding of the large differences between the results of the  various
sets, one should  use the ABKM and HERAPDF predictions as a reflection of the
theoretical  uncertainty in the game. This is very  important since, except from
MSTW, they are among the few other  parameterizations which are available at
NNLO, i.e.  the order required to address Higgs production at hadron colliders.
It is thus imperative  that one assesses the impact of using these  two sets for
the Higgs production cross section.  This is particularly important for a 
crucial issue such as the exclusion of the Higgs boson in a certain mass range.
\s

In conclusion, and in view of the above arguments, we strongly believe that the
analysis that we have developed in ours paper \cite{JHEP,PLB} scientifically
sound (see also Ref.~\cite{Sally}).  It could appear at first sight that we have been a little bit
conservative in the estimate of the theoretical uncertainties (although recent
analyses tend to show that it is far from being the case), but when it comes to
a such a crucial issue as excluding the Higgs boson (which we believe is the
most important issue in today high--energy physics), it is more recommended
than, to the opposite, being too aggressive. A too optimistic analysis that
excludes a possibility that can be discovered somewhere else, could affect
the credibility of our field. \bigskip

{\bf Acknowledgements}: We thank the organizers of the Rencontres de la Vall\'ee
d'Aoste as well as Moriond QCD and electroweak in La Thuile this winter for
their kind invitations to present our work and for making possible a critical
debate with the experimentalists on the issues presented here. We acknowledge
the projects SR/S2/JCB64 DST (India) and ANR CPV-LFV-LHC NT09-508531 (FR) for
support.

\subsection*{Note added}

Very recently, a new analysis of the PDF and $\alpha_s$ dependence of the Higgs
cross-section at the Tevatron and the LHC has been presented by members of the
MSTW PDF group \cite{Thorne:2011kq}. We would like to make here very brief
preliminary remarks:\s 

-- Our analyses in Refs.~\cite{JHEP,PLB} should, by no means, be considered as a
criticism of the work of the PDF fitters. The message that we wanted to convey
was  simple: there are other PDF sets than MSTW on the market and the spread 
between the  various predictions is much larger than the estimate of the PDF
uncertainty  made within one given parametrization. As already mentioned, we are
not PDF experts and we simply note that there is a problem (that
Ref.~\cite{Thorne:2011kq} also  acknowledges, although it is indeed less severe
than initially thought by us and the spread in the predictions is only at the
level of 25--30\% and not 40\%) which needs  to be addressed by the experts. 

-- The analysis of Ref.~\cite{Thorne:2011kq} concludes that the PDF sets which
do not take into account the Tevatron jet data fail to give a good description
of the same. However, it is to be noted that this differs from the conclusions
of Ref.~\cite{Alekhin:2011cf} which ascribes this as possibly being remedied 
once  the missing higher order corrections to the dijet calculations would be
included. It is to be hoped that the data from LHC may help in increasing our
understanding in the near future. But at present one can only conclude that the
issue is being debated by experts.

-- Ref.~\cite{Thorne:2011kq} does not recommend that one includes the
theoretical  uncertainty on $\alpha_s$ in addition to the PDF+$\alpha_s$
experimental uncertainty as we suggest in our studies. We insist again on the
fact this suggestion was made before the PDF4LHC recommendation \cite{PDF4LHC}
became available and that the results for the PDF+$\alpha_s$ uncertainty that we
obtain using our approach {\it are the same} as those  obtained using the
PDF4LHC recommendation.  

-- Ref.~\cite{Thorne:2011kq} focuses only on the PDF+$\alpha_s$ issue. This
is,   however, only part of the points that we raised in our studies. Indeed,
besides the  PDF+$\alpha_s$ problem, we have also addressed the issues with the
scale and with the EFT uncertainties as well as the combination of all sources
of  uncertainties. We reiterate our main conclusion that if the approach adopted
by the LHC Higgs cross section working group \cite{LHCXS} is applied to the
$gg\to H$  cross section at the Tevatron, one would find an uncertainty that is
much larger than that assumed  by the CDF/D0 collaborations (see again Fig.~1).
We therefore still believe that the CDF/D0 Higgs exclusion limits are not as 
robust as claimed.


\newpage

\begin{thebibliography}{999}

\bibitem{JHEP} J. Baglio and A. Djouadi, JHEP 1010 (2010) 064; arXiv:1003.4266
[hep-ph].\vspace*{-2mm}

\bibitem{PLB} J. Baglio, A. Djouadi, S. Ferrag and R. Godbole, Phys. Lett.  
B699 (2011) 368; arXiv:1101.1832 [hep-ph] and erratum to appear.\vspace*{-2mm} 

\bibitem{ggH-LO}  H. Georgi, S. Glashow, M. Machacek, D. Nanopoulos, Phys.
Rev. Lett. 40 (1978) 692.\vspace*{-2mm} 

\bibitem{ggH-NLO} A. Djouadi, M. Spira and P. Zerwas, Phys. Lett. B264 (1991)
440;   S. Dawson, Nucl. Phys. B359 (1991) 283.\vspace*{-2mm}

\bibitem{SDGZ}  M. Spira, A. Djouadi, D. Graudenz and P.M. Zerwas,  Nucl. Phys.
B453 (1995) 17.\vspace*{-2mm}

\bibitem{ggH-NNLO} R.V.~Harlander and W. Kilgore, Phys. Rev. Lett. 88 (2002)
201801; C. Anastasiou and K. Melnikov, Nucl. Phys. B646 (2002) 220; V. Ravindran,
 J. Smith and W.L. Van Neerven, Nucl. Phys. B665 (2003) 325.\vspace*{-2mm}  

\bibitem{ggH-resum} S.~Catani, D.~de Florian, M.~Grazzini and P.~Nason, JHEP
0307 (2003) 028.\vspace*{-2mm}

 
\bibitem{ggH-FG} D. de Florian and M. Grazzini, Phys. Lett. B674 (2009) 291.\vspace*{-2mm}

\bibitem{ggH-radja} C. Anastasiou, R. Boughezal and F. Pietriello, 
JHEP 0904 (2009) 003.\vspace*{-2mm}


\bibitem{ggH-EW} S. Actis, G. Passarino, C. Sturm and S.  Uccirati, Nucl. Phys.
B811 (2009) 182.\vspace*{-2mm}

\bibitem{HV-all} S. Glashow, D. Nanopoulos and A. Yildiz, Phys. Rev. D18
(1978) 1724; T. Han and S. Willenbrock, Phys. Lett. B273 (1991) 167;  A. Djouadi
and M. Spira, Phys. Rev. D62 (2000) 014004;   O. Brein, A. Djouadi and
R. Harlander, Phys. Lett. B579 (2004) 149.\vspace*{-2mm}  

\bibitem{Tevatron0} The CDF and D0 collaborations,   Phys. Rev. Lett. 104 (2010)
061802; updated in  arXiv:1007.4587 [hep-ex].\vspace*{-2mm} 

\bibitem{Tevatron1} The CDF and D0 collaborations, arXiv:1007.4587 [hep-ex].\vspace*{-2mm}

\bibitem{CDF} The CDF collaboration, CDF note 10232, 16/08/2010 (the note is apparently not
accessible anymore directly from the CDF server).\vspace*{-2mm} 


\bibitem{Response1} 
{\small http://tevnphwg.fnal.gov/results/SMHPubWinter2010/gghtheoryreplies\_may2010.html}\vspace*{-2mm}

\bibitem{Response2}  {\small
 http://tevnphwg.fnal.gov/results/SM\_Higgs\_Summer\_10/addendumresponse\_oct2010.html}\vspace*{-2mm}

\bibitem{Thuile} M. Buehler, on behalf of the CDF/D0 collaborations and
TEVNPHWG, talk given at the  XXV$^{\rm emes}$ Rencontres de la Vall\'ee d'Aoste,
La Thuile, March 2011; {\small
http://agenda.infn.it/conferenceOtherViews.py?view=standard\&confId=3080}.  See
also, the talks of K. Petridis and B. Jayatilaka at Moriond (electroweak) 
2011.\vspace*{-2mm} 

\bibitem{Moriond} J. Hayes on behalf of the CDF and D0 collaborations, 
Combinations of Searches for SM Higgs at the Tevatron, talk given at the 
Moriond QCD 2011 conference, March 2011,  {\small
http://moriond.in2p3.fr/QCD/2011/MorQCD11Prog.html}\vspace*{-2mm}

\bibitem{Moriond-AD} A. Djouadi, talk given at the  Moriond QCD 2011 conference,
March 2011,  {\small http://moriond.in2p3.fr/QCD/2011/MorQCD11Prog.html}\vspace*{-2mm}

\bibitem{LHCXS} S. Dittmaier et al., ``Handbook of LHC Higgs cross sections", 
arXiv:1101.0593 [hep-ph].\vspace*{-2mm} 


\bibitem{PDF-HERA}  See {\tt
www.desy.de/h1zeus/combined\_results}.\vspace*{-2mm}

\bibitem{PDF-ABKM}  S. Alekhin et al.,  Phys. Rev. D81 (2010)
014032.\vspace*{-2mm} 


\bibitem{PDF-MSTW} A. Martin, W. Strirling, R. Thorne and G. Watt,  Eur. Phys.
J. C63 (2009) 189.\vspace*{-2mm}


\bibitem{PDF-MSTW1}  A. Martin, W. Strirling, R. Thorne and G. Watt,  Eur. Phys.
J. C64 (2009) 653.\vspace*{-2mm}


\bibitem{ADGSW} C. Anastasiou et al., JHEP 0908 (2009) 099.\vspace*{-2mm}

\bibitem{Scale-H0j} C.F.~Berger, C.~Marcantonini, I.W.~Stewart, F.J.~Tackmann 
and W.~J.~Waalewijn, JHEP 1104  092 (2011).\vspace*{-2mm}


\bibitem{LHC-jet} See the summary given by R. Tanaka at the LHC Higgs Combination
Group open meeting (May 2011), {\small
http://indico.cern.ch/conferenceDisplay.py?confId=135332}.\vspace*{-2mm}


\bibitem{ggH-Ahrens} V. Ahrens et al., Eur. Phys. J. C62 (2009) 333.\vspace*{-2mm}


\bibitem{Paolo}  A. Djouadi and P. Gambino, Phys. Rev. Lett. 73 (1994) 2528; 
A. Djouadi, P. Gambino and B. Kniehl, Nucl. Phys. B523 (1998) 17.\vspace*{-2mm}

\bibitem{Hoogeven}
J.J. van der Bij and F. Hoogeveen, Nucl. Phys. B283 (1987) 477.\vspace*{-2mm} 

\bibitem{Awramik}
M. Awramik, M. Czakon, A. Freitas, G. Weiglein, Phys. Rev. Lett. 93 
(2004) 201805.\vspace*{-2mm} 


\bibitem{PDF4LHC}  M. Botje et al.,  e-Print: arXiv:1101.0538 [hep-ph].\vspace*{-2mm}  


\bibitem{PDF-CTEQ} P.M. Nadolsky et al. (CTEQ coll.), Phys. Rev. D78 (2008) 
013004.\vspace*{-2mm}  

\bibitem{PDF-NN} R.D. Ball et al., Nucl. Phys. B823 (2009) 195.\vspace*{-2mm}  

\bibitem{PDF-ttpaper}  M. Cacciari, S. Frixione, M. Mangano, P. Nason, G.
Ridolfi, JHEP 0809 (2008) 127.\vspace*{-2mm}

\bibitem{Sally} Sally Dawson, talk given at the University of Washington 
Workkshop on Higgs cross section ``Higgs @ Tevatron and LHC", April 2011,
where 
{\small
http://silicon.phys.washington.edu/Higgs-at-Tevatron-and-LHC-Talks/Sally\_Dawson.pdf}.



\bibitem{NMC} S. Alekhin, J. Blumlein and S. Moch, arXiv:1101.5261.\vspace*{-2mm} 


\bibitem{ABM10} See also, S. Alekhin, talk  at INT Seattle workshop. November 2010
(pages 11 and 12) 
{\small http://www.int.washington.edu/talks/WorkShops/int\_10\_3/People/Alekhin\_S/Alekhin.pdf};
S. Moch, talk  at Kick-off meeting of the LHCPhenoNet,  Feb. 2011,  Valencia,
(pp 8-9) {\small
http://indico.ific.uv.es/indico/contributionDisplay.py?contribId=5\&sessionId=2\&confId=339}.


\bibitem{PDF-all} S. Alekhin et al., arXiv:1011.6259 [hep-ph]. \vspace*{-2mm}

\bibitem{Voica} V. Radescu, talk given at the  Moriond QCD 2011 conference,
March 2011,  {\small
http://moriond.in2p3.fr/QCD/2011/MorQCD11Prog.html}\vspace*{-2mm}

\bibitem{JHEP-LHC} J. Baglio and A. Djouadi, JHEP 1103 (2011) 055; 
arXiv:1012.0530 [hep-ph].\vspace*{-2mm}


\bibitem{MSSM} J. Baglio and A. Djouadi, Phys. Lett. B699 (2011) 372; 
arXiv:1103.6247 [hep-ph].\vspace*{-2mm}



\bibitem{Thorne:2011kq} R.S.~Thorne and G.~Watt, ``PDF dependence of Higgs cross
 sections at the Tevatron and LHC: response to  recent criticism,''
 arXiv:1106.5789 [hep-ph].\vspace*{-2mm}


\bibitem{Alekhin:2011cf} S.~Alekhin, J.~Blumlein and S.-O.~Moch, ``Parton
distributions and Tevatron jet data,'' arXiv:1105.5349 [hep-ph]. 


\end{thebibliography}
\end{document}